# Idealized Slab Plasma approach for the study of Warm Dense Matter

A. Ng[1,2], T. Ao[2], F. Perrot[3], M.W.C. Dharma-Wardana[4] and M.E. Foord[1]

[1] Lawrence Livermore National Laboratory, Livermore, CA, U.S.A.

[2] Dept. of Physics & Astronomy, Univ. of British Columbia, Vancouver, B.C., Canada

[3] CEA Bruyeres Le Chatel Cedex, France

[4] IMS, National Research Council, Ottawa, Ontario, Canada, K1A 0R6

Recently, Warm Dense Matter has emerged as an interdisciplinary field that draws increasing interest in plasma physics, condensed matter physics, high pressure science, astrophysics, inertial confinement fusion, as well as material science under extreme conditions.  To allow the study of well-defined Warm Dense Matter states, we have introduced the concept of Idealized Slab Plasma that can be realized in the laboratory via (i) the isochoric heating of a solid and (ii) the propagation of a shock wave in a solid.  The application of this concept provides new means for probing AC conductivity, equation of state, ionization and opacity.  These approaches are presented here using results derived from first-principles theory and numerical simulations.

## I. Introduction

Warm Dense Matter (WDM) refers to states with comparable thermal and Fermi energies and ion-ion coupling parameters [1] that exceed unity. This classification was introduced to draw attention to a class of states that is relevant to a wide range of disciplines [2]. A particularly interesting aspect of WDM is that it has a different identity in different disciplines. Figure 1(a) shows a phase diagram of aluminum described by the Quotidian Equation of State (QEOS) model [3]. The line $kT=E_F$ is the loci of states with equal thermal ($kT$) and Fermi ($E_F$) energy. $G_{ii}$ is the ion-ion coupling parameter which is the ratio of the Coulomb potential between two neighboring ions to their thermal energy in accordance with the one component plasma description [1]. While there is no exact boundaries for the WDM regime, for convenience it is generally taken to be the region of the phase diagram where temperatures range from 1 to 100 eV and densities from 0.1 to 10 times solid density. In the context of Figure 1, WDM appears as strongly coupled plasmas characterized by $G_{ii}>1$. However in the context of Figure 1(b) WDM is then seen as high-temperature condensed matter or as materials under extreme conditions of high pressure. WDM also plays an important role in astrophysics and inertial confinement fusion. A classic example is deuterium [4,5] that is shock compressed to conditions pertinent to the outer core of Jupiter [6] and fusion capsules [7]. Other examples of the WDM regime can be found elsewhere [8].

Fig. 1

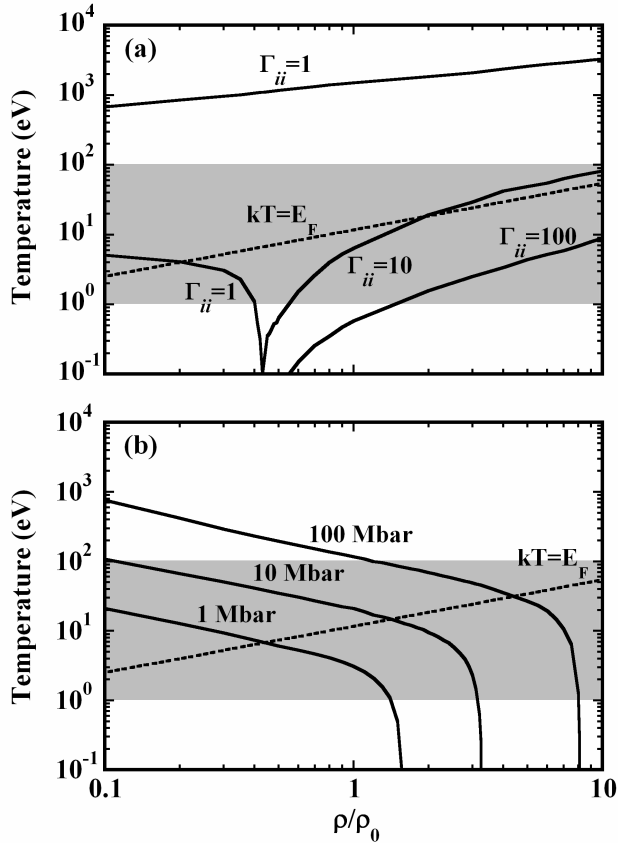

**WDM as (a) strongly coupled plasmas and (b) high-temperature condensed matter or high pressure states. *r* is mass density and *r<sub>o</sub>* is solid density of aluminum at normal conditions. The shaded area provides a convenient indication of the WDM regime.**

Understanding WDM is key to understanding the convergence between condensed matter physics and plasma physics, which has long been a difficult regime because of the difficulty of first-principles theories and scarcity of well defined experimental measurements. This can be best appreciated from the widely used Sesame tabulated equation of state, a *tour de force* theoretical effort to provide global physical data over a wide range of densities and temperatures [9]. In spite of the elaborate combination of seven distinct models ranging from semi-empirical to first-principle, the WDM regime is described only by interpolations between models.

Theoretical studies of WDM are challenging. Viewed as a strongly coupled plasma dominated by ion-ion correlation, WDM cannot be treated with the conventional Debye screening and perturbative approaches. Alternatively, when viewed as high temperature condensed matter, it is a disordered system whose description requires detailed knowledge of excited states, structure factors and the dynamics of strongly interacting electrons and ions. Experimental studies of WDM are equally daunting. This stems from the extreme pressure associated with such states (Figure 1(b)). Their production requires sources that can deliver high energy densities. An even greater challenge is the measurement of physical properties of uniform, well defined states for unambiguous comparisons with theory when no physical means is available to confine such states.

For laboratory investigations the ideal configuration would be a planar sample with uniform density and temperature, whose state can be characterized by model independent measurements. Any measured properties can then be attributed to a single state, thus allowing the direct test of theory in a well defined manner. To realize this in the laboratory, we have introduced the concept of Idealized Slab Plasma (ISP), a high-density plasma in planar geometry that can be considered a uniform slab of WDM in so far as all residual non-uniformities have insignificant impact on the measurement of its properties. In this paper, we describe two approaches of this concept based on (i) isochoric heating and (ii) shock compression of a solid. We also give examples of how these can be used to obtain well defined, single-state measurements.

## II. Idealized Slab Plasma produced by isochoric heating of a solid

This approach is based on energy deposition in a solid by an ultra-fast source. By limiting the heating process to a femto-second time scale, hydrodynamic expansion is minimized and isochoric conditions can be produced. At the same time, uniform heating is ensured by matching the thickness of the sample to the deposition range of the energy source as well as the characteristic scale-length for thermal conduction. The latter is determined by the mean-free-path of the thermal electrons. It is quite remarkable that the thickness of such an Idealized Slab Plasma (ISP) can be tailored to vary over six orders of magnitude by using different energy sources as suggested by the *1/e* absorption lengths in aluminum for optical photons, X-rays, electrons and ions (Figure 2). Even thicker systems can be attained with the use of heavy ions. Thus, this approach is widely scalable in term of the thickness of the sample studied.

To characterize an ISP produced by isochoric heating of a solid, one can use its mass density and excitation energy density. The former is simply the initial density of the solid since expansion of the sample is negligible. The latter is determined by the absorbed energy density imparted by the heating process, which far exceeds the energy density of the unperturbed solid.

**Fig. 2**

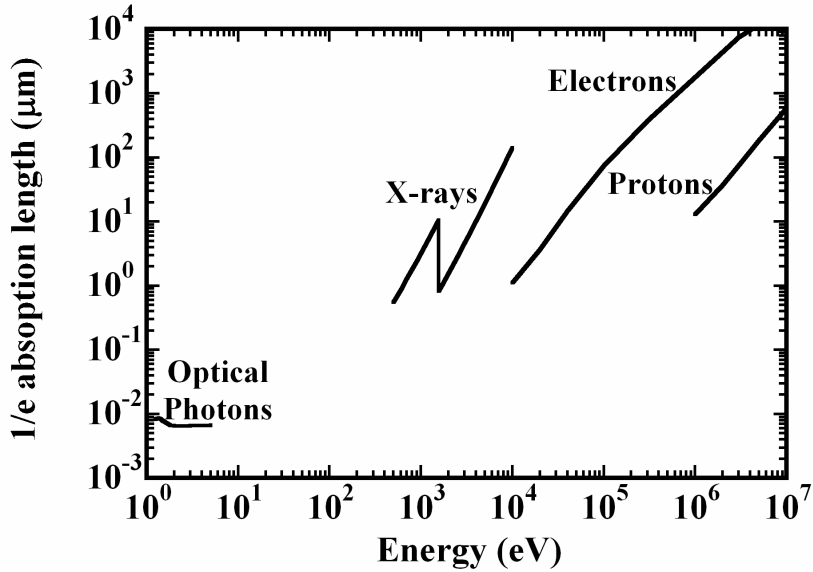

*1/e* **absorption lengths in aluminum for lasers, x-rays, electrons and protons.**

II.a    Studying the AC conductivity of WDM.

Among the various ultra-fast energy sources, femtosecond lasers are the most developed and accessible. The first ISP concept that has emerged is the isochoric heating of an ultrathin foil by a femtosecond laser [10]. This approach was introduced using the example of a 100Å aluminum foil heated by a pump laser pulse of 20fs (FWHM) at a wavelength of 400nm. However, it is important to recognize that the concept can be readily extended to lasers with much longer pulse length and foils with significantly greater thickness by choosing the appropriate pump laser intensity. To demonstrate this versatility, we consider here the case of a 400nm, 100fs laser pulse incident normally on a 200Å thick aluminum foil. These requirements can readily be met in the laboratory.

The numerical simulations used in this work are based on a 1-dimensional hydrodynamic code [11] in which the laser-heated foil is considered a two-temperature, electron and ion fluid. For this preliminary study, the equation of state of this plasma fluid is given by QEOS [2] that is derived from an averaged atom, screened hydrogenic model where the electrons are treated as a Thomas Fermi gas and the ions a Cowan fluid. Consistent with this Thomas-Fermi type EOS, the conductivities are obtained from the dense plasma model of Lee and More [12]. This is based on the Boltzmann transport equation in the relaxation time approximation. Both electron-ion and electron-neutral scattering are included. Coulomb interaction is described by a modified cross-section with cutoffs in the strong coupling limit. In order to avoid the complexities of a wide-range calculation of the energy equilibration rate between electrons and ions, we have adopted the use of a phenomenological coupling constant $g$ [13]. The value of $g$ for aluminum is taken to be $10^{17}$ W/m$^3$K in accordance with that deduced from the measurement of electron temperature at a shock front [14]. We have also assumed that hydrodynamic expansion of the heated solid occurs when the ion (lattice) temperature reaches the melting point.

The interaction of the heating or pump laser pulse with the foil is evaluated from the Helmholtz equations for an electromagnetic wave of S or P polarization. The resulting plasma fluid is treated as a dielectric medium whose dielectric function is derived from the Drude model in which the DC conductivity is determined by the collision frequency [12]. A detailed description of this electromagnetic wave solver has been given elsewhere [15].

Figure 3 shows the calculated reflectivity, transmission and absorption of the pump laser pulse by the foil. The spatial profiles of energy density (including kinetic energy), mass density and electron temperature in the foil at the end of the pump pulse are presented in Figure 4. The heated foil appears to maintain a slab-like density structure with nearly uniform energy density and temperature in the slab region. The

**Fig. 3**

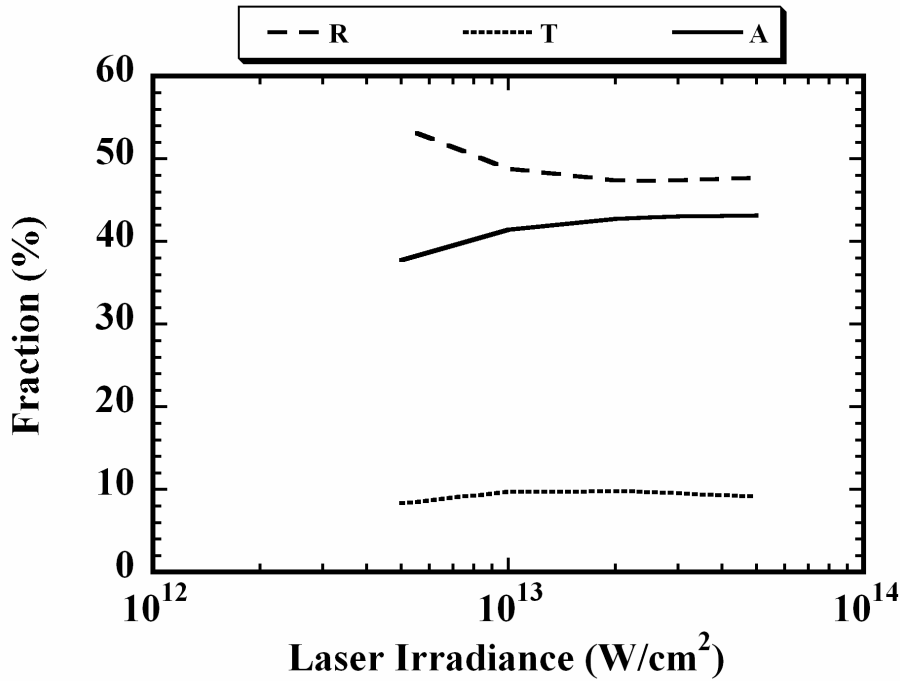

**Calculated reflectivity, transmission and absorption as a function of laser irradiance for a 400nm, 100fs pump laser pulse incident normally onto a 200Å aluminum foil.**

uniformity in energy density and temperature is the result of electron thermal conduction. For the conditions of interest, the Fermi energy is about 12.5 eV and the Fermi speed is about $2 \times 10^8$ cm/s. Accordingly the transit time of the electrons near the Fermi surface across a 200Å foil is ~10 fs. Thus the absorbed laser energy can be distributed nearly

**Fig. 4**

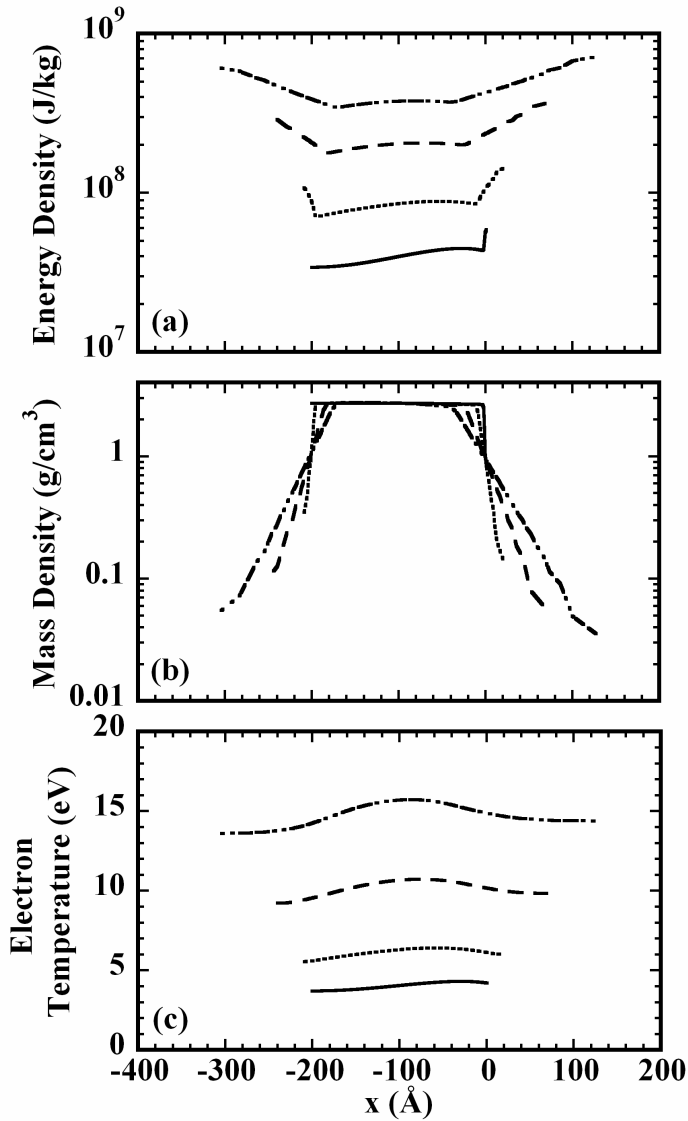

(a) Spatial profiles of energy density, (b) mass density and (c) electron temperature at the end of the pump pulse for irradiances of $5 \times 10^{12}$ W/cm$^2$ (solid line), $10^{13}$ W/cm$^2$ (dotted line), $2.5 \times 10^{13}$ W/cm$^2$ (dashed line) and $5 \times 10^{12}$ W/cm$^2$ (dot-dashed line). The position of x=0 corresponds to the location of the front (laser facing) surface of the foil.

uniformly in the sample during the heating process. At the lowest irradiance, virtually no hydrodynamic expansion has occurred at the end of the pulse since only a very thin layer on the front surface has an ion temperature reaching the melting point of aluminum.

Expansion in the outer surfaces of the foil becomes noticeable at irradiances of $2.5 \times 10^{13}$ W/cm$^2$ and above.

It should be noted that in experiments the expansion of the foil can be monitored using a diagnostic such as frequency domain interferometry in a pump and probe arrangement [16, 17]. This measures the change in phase shift of a femtosecond probe laser pulse reflected from the surface of the foil. The response of such a diagnostic can be predicted using the electromagnetic wave solver for the probe pulse in a post-processor calculation based on results of the hydrodynamic simulation. Results of some sample calculations are presented in Figure 5. Zero and 300 fs correspond respectively to the times of peak pump laser intensity and end of the pump pulse. The initial change in phase shift is due to change in the AC conductivity of the heated foil. For the higher irradiances hydrodynamic expansion becomes evident at late times, appearing as a continual decrease in the phase shift of the reflected probe.

**Fig. 5**

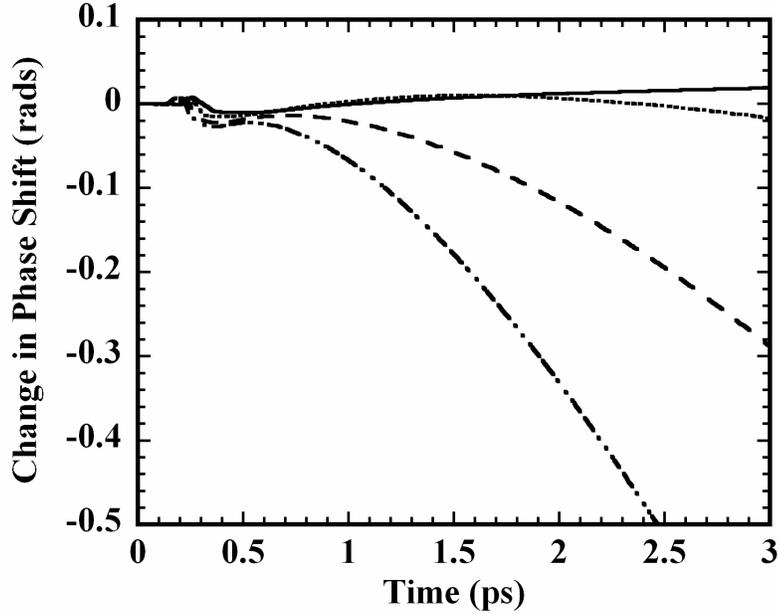

**Change in phase of a 100fs, 400nm, S-polarized, 45°-incident probe reflected from the front surface of the foil target for pump laser irradiances of $5\times10^{12}$ W/cm$^2$ (solid line), $10^{13}$ W/cm$^2$ (dotted line), $2.5\times10^{13}$ W/cm$^2$ (dashed line) and $5\times10^{13}$ W/cm$^2$ (dot-dashed line).**

Because of the slab-like behavior, the mass averaged density of the foil at the end of the pump laser pulse remains close to its initial density ($r_o$) as illustrated in Figure 6(b). Furthermore, because of the comparatively low initial energy density of the foil and the near uniform deposition of the pump laser, the absorbed or excitation energy density (*De*) provides a very good measure of the mass averaged energy density at the end of the pump pulse as shown in Figure 6(a). The state of the slab plasma can thus be characterized with $r_o$ and *De*. However, this does not constitute a complete description of

Fig. 6

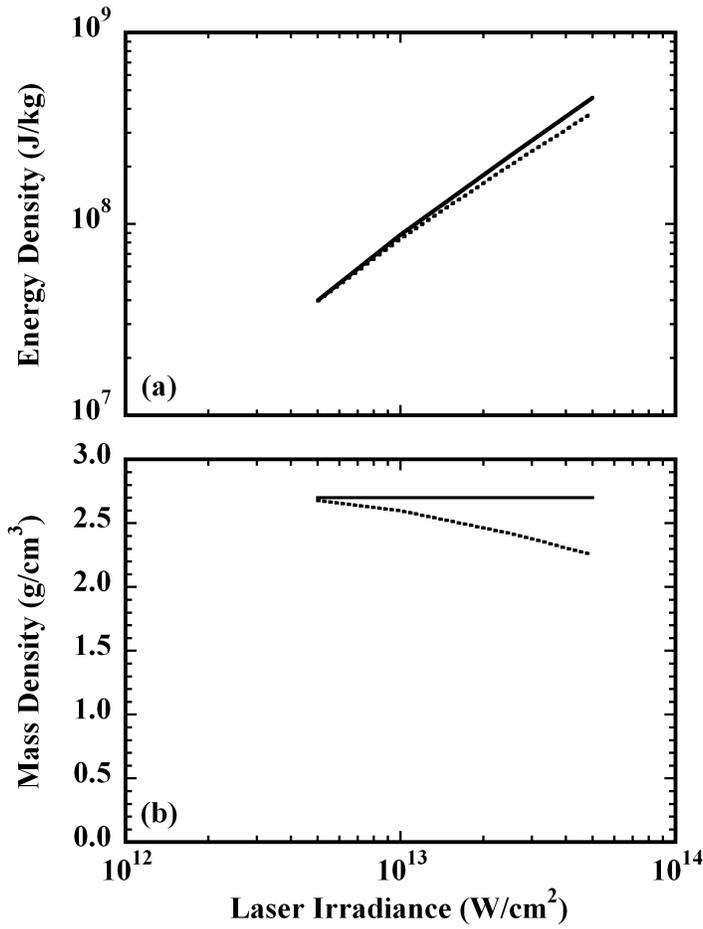

**(a) Comparison of the absorbed energy density (solid line) to the mass averaged energy density (dotted line) and (b) comparison of the initial solid density (solid line) to the mass averaged mass density (dotted line) as a function of pump laser irradiance.**

the slab since the electron and ion temperatures are different as dictated by the finite rate of energy exchange between the two species. Figure 7(a) shows the mass averaged electron and ion temperatures for $g=10^{17}$ W/m$^3$K [14]. The corresponding contributions of the electrons and ions to the total energy density are shown in Figure 7(b). This indicates that the ion contribution is largely negligible in the regime of interest here. It is also evident in Figure 7(a) that the ion temperature is less than 5% of the electron temperature even for the highest irradiance of interest here.

**Fig. 7**

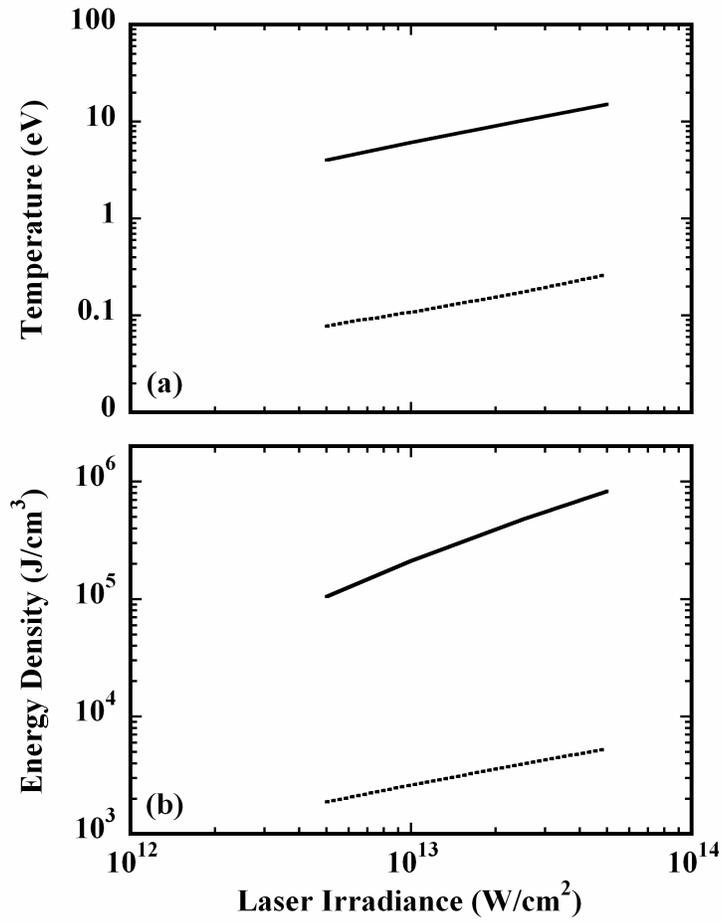

**(a) Mass averaged electron (solid line) and ion (dotted line) temperatures and (b) mass averaged energy densities of the electrons (solid line) and ions (dotted line) as a function of pump laser irradiance.**

Next, we consider the use of the femtosecond-laser heated ultrathin foil as a means to study the AC conductivity, $\sigma(\omega)$, of WDM based on measurements of reflectivity ($R$) and transmission ($T$) of the heated foil. The conventional, forecasting method is to compare the calculated values of $R$ and $T$ with their measured values to test the conductivity model used in the hydrodynamic simulation and the electromagnetic

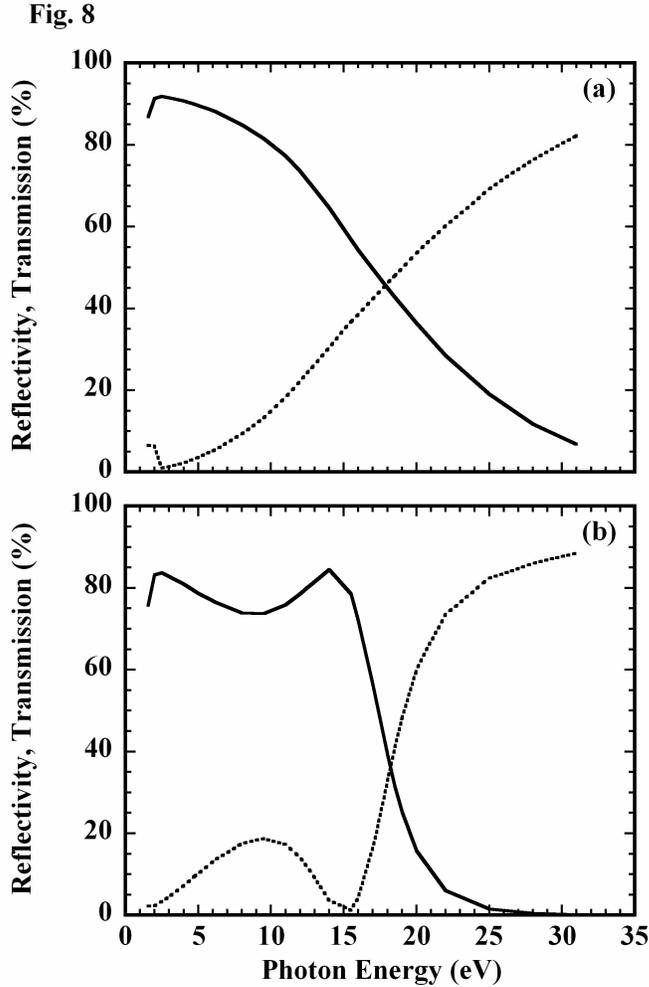

**Fig. 8**

**Calculated reflectivity (solid line) and transmission (dotted line) of an unperturbed 200Å aluminum foil as a function of photon energy for a 45°-incident (a) S-polarized and (b) P-polarized probe.**

wave solver. As an example, we consider a probe pulse of 100fs (FWHM) incident on the front surface of the foil at an angle of 45°. To probe AC conductivity over a broad range of frequencies, one can use high harmonics generated from monomer gases [18, 19], molecules [20], solids [21,22] or clusters [23] heated with femtosecond lasers, or femtosecond pulses from a synchrotron light source. The calculated reflectivity and transmission of an unperturbed foil as a function of photon energies for S and P polarized probes are shown in Figures 8(a) and 8(b) respectively. The corresponding reflectivity $R$ and transmission $T$ for the heated foil at different pump laser irradiances are shown in Figures 9 and 10. The hydrodynamic expansion of the foil within the duration of the probe pulse is taken into account in these calculations. Also included in Figures 9 and 10 are values of $R_{Slab}$ and $T_{Slab}$ that are evaluated by treating the foil as an idealized slab

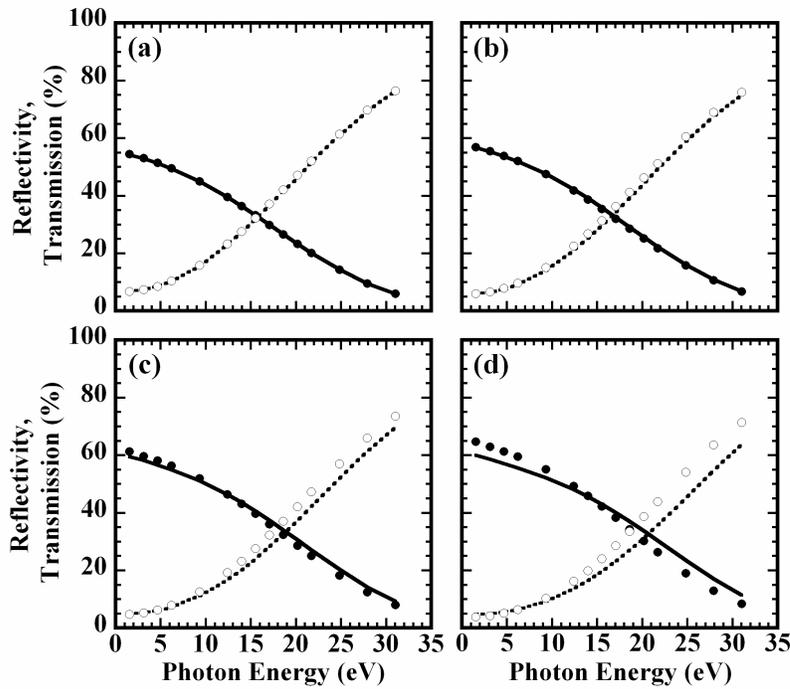

**Fig. 9**

**Calculated reflectivity ($R$ and $R_{Slab}$) and transmission ($T$ and $T_{Slab}$) as a function of photon energy of a S-polarized probe at 45° angle of incidence for the heated foils at pump irradiances of (a) $5 \times 10^{12}$ W/cm$^2$, (b) $10^{13}$ W/cm$^2$, (c) $2.5 \times 10^{13}$ W/cm$^2$, and (d) $5 \times 10^{13}$ W/cm$^2$. Values of $R$ (solid circle) and $T$ (open circles) are obtained from calculations taking into expansion during the heating pulse. Values of $R_{Slab}$ (solid line) and $T_{Slab}$ (dotted line) are for the idealized slab.**

specified by $r_o$ and $D_e$. We have approximated the ion temperature as 5% of the electron temperature as discussed above. The convergence between $\{R, T\}$ and $\{R_{Slab}, T_{Slab}\}$ is a measure of the validity of the Idealized Slab Plasma approach. It is evident from Figure 9 and 10 that S-polarized observations are much less sensitive to hydrodynamic expansion due to the absence of resonant absorption.

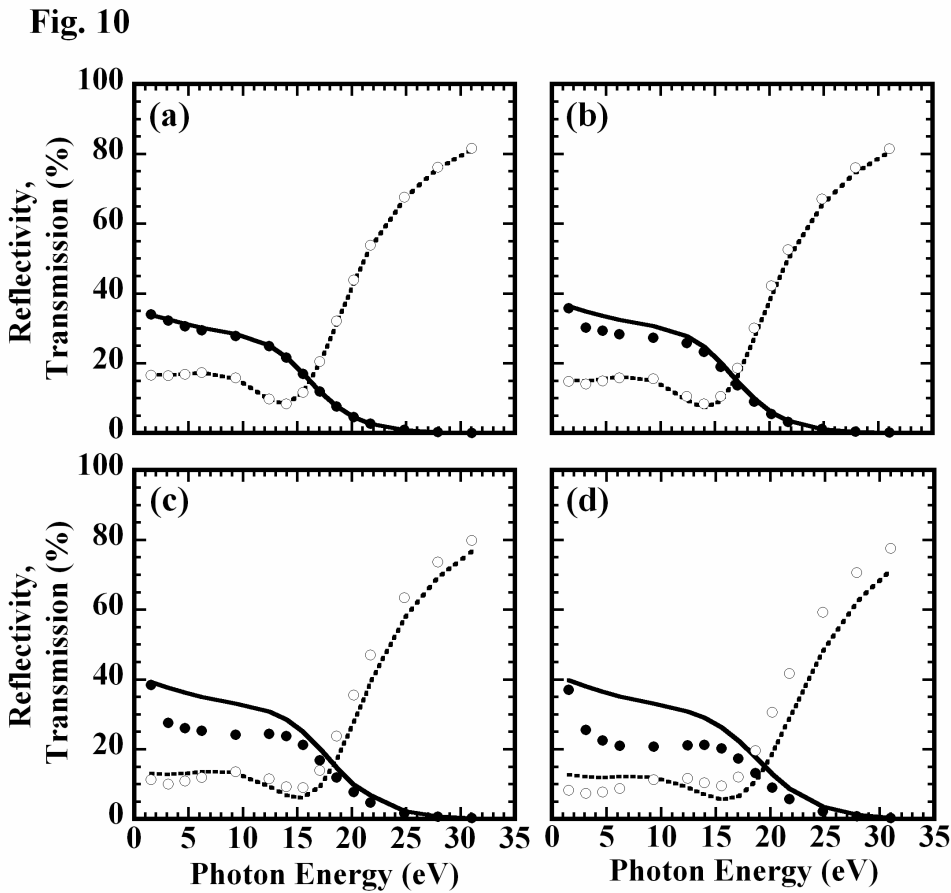

**Fig. 10**

Calculated reflectivity ($R$ and $R_{Slab}$) and transmission ($T$ and $T_{Slab}$) as a function of photon energy of a P-polarized probe at 45° angle of incidence for the heated foils at pump irradiances of (a) $5\times10^{12}$ W/cm$^2$, (b) $10^{13}$ W/cm$^2$, (c) $2.5\times10^{13}$ W/cm$^2$, and (d) $5\times10^{13}$ W/cm$^2$. Values of $R$ (solid circle) and $T$ (open circles) are obtained from calculations taking into expansion during the heating pulse. Values of $R_{Slab}$ (solid line) and $T_{Slab}$ (dotted line) are for the idealized slab.

The more novel methodology for studying $\sigma(\omega)$ is the backcasting approach of using the observed values of $\{R, T\}$ to solve the Helmholtz equations for electromagnetic wave propagation in a plasma slab that is characterized by $r_o$ and $D_e$. The solution yields the dielectric function $\varepsilon_\omega$ or the AC conductivity $\sigma_\omega$ (both real and imaginary parts) of the plasma independent of any simulation models. The value of $s_w(r_o, De)$ can then be used as a direct test of conductivity models for a single, well-defined plasma state. A detailed description of this methodology has been given earlier for the case of a single-frequency optical probe [10]. The use of high harmonics probes will allow a much more complete test of conductivity models over a broad frequency range.

II.b    Non-equilibrium equation of state of WDM

As noted above, a unique advantage of the isochoric heating approach is the scalability in sample thickness with appropriate ultrafast energy sources. Potential candidates include $K\alpha$ X-ray lines, energetic electrons and fast protons produced in strong-field laser-plasma interactions, or X-ray free electron lasers. As an example, we consider the heating of a 5000-Å thick aluminum foil by a femtosecond Si $K\alpha$ pulse to produce near uniform deposition and isochoric heating with an energy density of $3 \times 10^7$ J/kg. This yields an Idealized Slab Plasma with an electron temperature of about $4 \times 10^4$ K. The required $K\alpha$ radiation can be readily generated by the irradiation of a silicon slab with a femtosecond laser pulse at an irradiance of $\sim 10^{17}$ W/cm$^2$. Hot electrons accelerated by the ponderomotive potential of the laser light induce K-shell excitation in the cold region of the target. This gives rise to $K\alpha$ line emission as the K-shell vacancies are filled.

The significance of such a relatively thick Idealized Slab Plasma is its opacity to its characteristic emission spectrum. This renders it possible to use brightness temperature as a measure of electron temperature. In experiments, one can determine simultaneously the absorbance (or emissivity) and emittance (or spectral radiance) of the plasma slab. The former is obtained from reflectivity measurement. For the above example, the calculated reflectivity and emittance at 500 nm for the slab plasma convoluted with Gaussian temporal functions with Full-Widths at Half-Maximum of 500fs, 1 ps and 2 ps are presented respectively in Figures 11(a) and 11(b). These

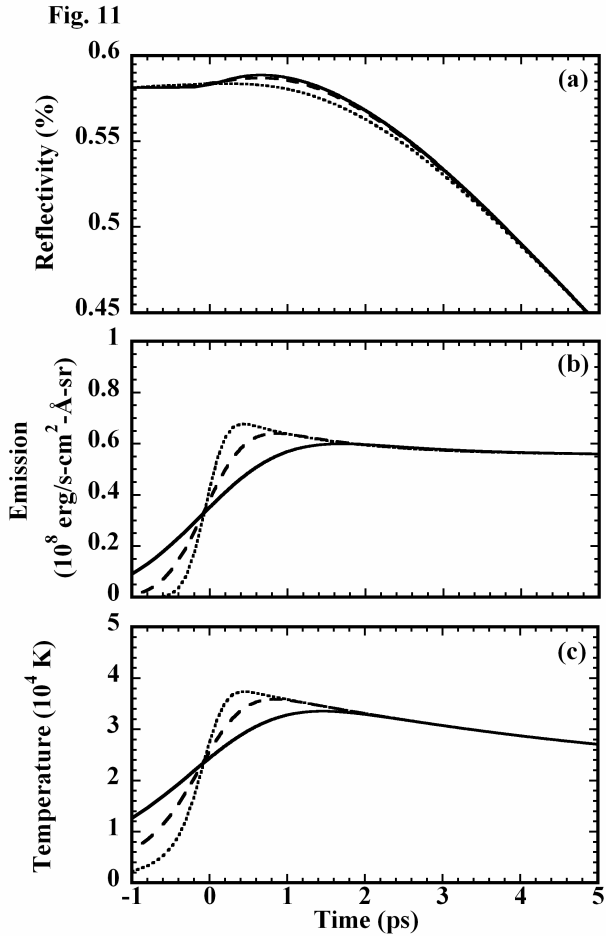

**Calculated (a) reflectivity, (b) emittance and (c) brightness temperature $T_{br}$ at 500nm as a function of time of a 5000Å thick, solid-density idealized slab plasma with an excitation energy density of $3 \times 10^7$ J/kg for temporal resolutions of 500fs (solid line), 1 ps (dashed line) and 2 ps (dotted line).**

simulate experimental measurements with temporal resolutions of 500 fs, 1 ps and 2 ps respectively. The results are used in Kirchoff's law to yield the brightness temperature $T_{br}$ (Figure 11(c)). Figure 12 shows the ratio of the peak value of $T_{br}$ to the electron temperature $T_{Slab}$ of the Idealized Slab Plasma as determined by its energy density, assuming that the ion temperature reaches 5% of the electron temperature. Almost identical values are obtained for calculations at 400 nm and 600 nm. The results suggest

**Fig. 12**

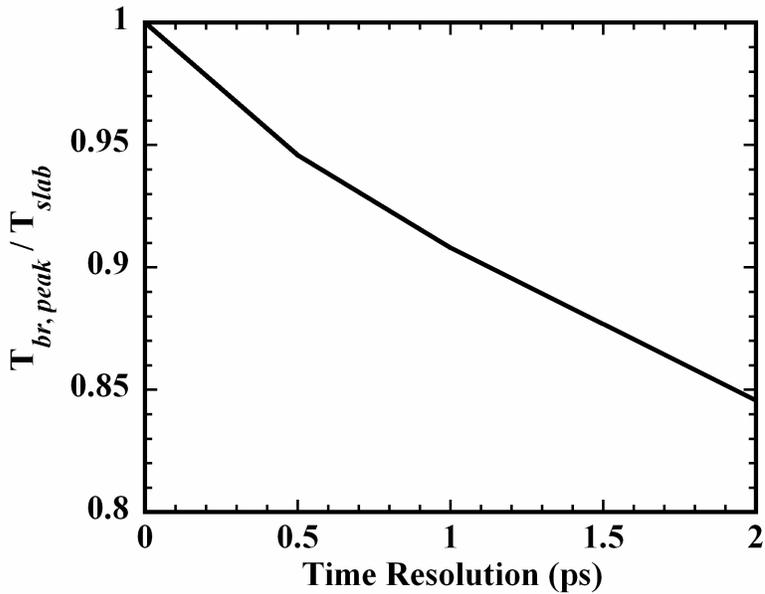

**Ratio of the peak brightness temperature $T_{br}$ to the electron temperature $T_{Slab}$ of the idealized slab as a function of the temporal resolution of observations at 500 nm.**

that for observations with a temporal resolution of 500 fs, $T_{br}$ provides a measure of $T_{Slab}$ to within 94% of its value. Such a temporal resolution can be attained with state-of-the-art streak cameras, for example, the Hamamatsu FESCA-200 single-shot streak camera that combines a temporal resolution of 200fs with a dynamic range better than 1:40. Thus, it appears feasible to obtain a reasonable measurement of the non-equilibrium equation of state of such WDM by determining electron temperature as a function of energy and mass densities, using this Idealized Slab Plasma approach. It should be noted such an experiment could be performed using the 100TW, 100fs Ti:Sapphire laser (JanUSP) in the Lawrence Livermore National Laboratory.

## III. Idealized Slab Plasma produced by a shock wave in solid

One of the most powerful techniques for producing WDM state in the laboratory is the propagation in a solid of a single, steady shock wave that can be readily generated with intense lasers, X-ray or particle beams. For such a shock wave passing through an initially unperturbed medium, the resulting state is constrained uniquely by the principal Hugoniot (Figure 13) which arises from the conservation of mass, momentum and energy across the shock front. [24].

Fig. 13

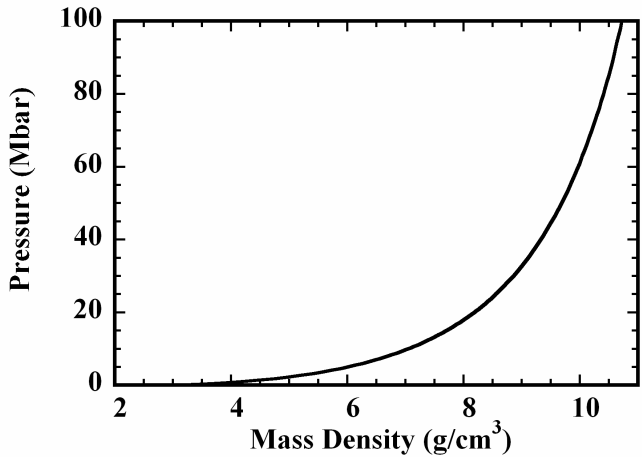

**Principle Hugoniot of aluminum described by QEOS [2].**

To illustrate the use of shock waves to produce a laboratory ISP, we consider a steady shock launched from a pusher layer of silicon into a sample layer of aluminum. Silicon is chosen as the pusher material because of its close match to aluminum in shock impedance. It is also an appropriate X-ray window in the spectral region of the K-shell photo-absorption lines and edge of aluminum. The shock pressure in aluminum is taken to be 25 Mbar. Figure 14 shows snapshots of the mass density, electron temperature and ion temperature profiles as calculated from our hydrodynamic code. The values of *g* are $10^{16}$ W/m$^3$K for silicon [13] and $10^{17}$ W/m$^3$K for aluminum [14]. Thermal gradients

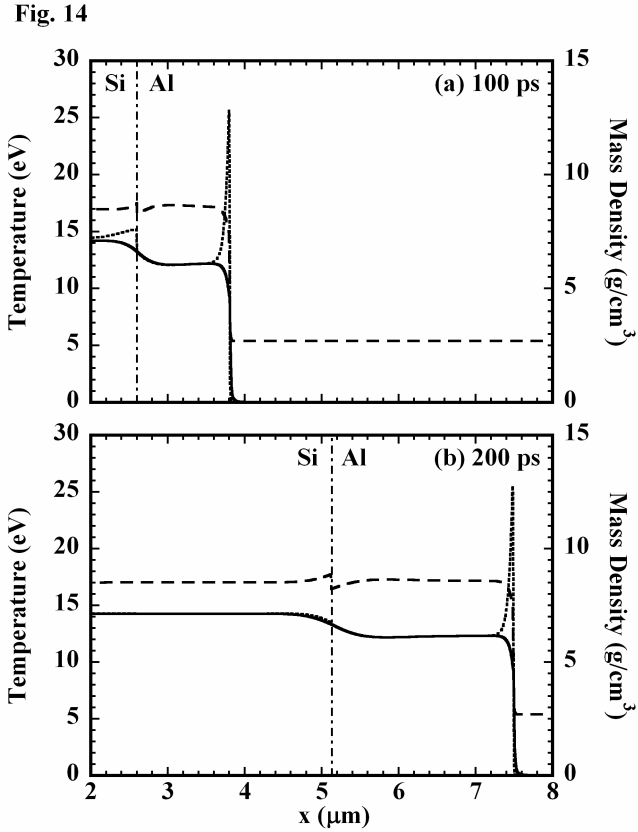

**Fig. 14**

**Snapshots of the mass density (dashed line), electron temperature (solid line) and ion temperature (dotted line) profiles of a shock wave propagating from silicon into aluminum. Shock pressure in aluminum is 25 Mbar. Time zero corresponds to the time of shock arrival at the silicon-aluminum interface.**

appear at both the silicon-aluminum interface and the shock front in aluminum. The former is due to differences in the equation of state of the materials and the latter is a manifestation of the electron-ion equilibration process. However, the main body of the shock compressed aluminum is completely uniform. Furthermore, the length of the shocked aluminum region grows linearly with time, mitigating rapidly the effect of the aforementioned thermal gradients and leading to the formation of an ISP whose state can be determined uniquely by measuring the shock speed $U_S$. It is evident that implicit in this approach is the scalability of the thickness of the sample.

III.a    Pressure ionization of WDM

Ionization physics is fundamental to the determination of all basic properties of WDM. The macroscopic parameter of average ionization, <Z>, is central to some transport models because of the need to evaluate electron density for the calculation of transport coefficients. Although a variety of ionization models exist, there is substantial divergence in their predictions even for <Z> as noted earlier [25]. This is further illustrated in Figure 15 by comparing earlier values of <Z> obtained from QEOS [2], Sesame tabulated equation of state [9] and a detailed configuration model [25] to those derived from a neutral-psuedo-atom density-functional-theory (DFT) model [25] for aluminum at 12.5 eV. Such discrepancies remain unresolved due to the lack of experimental data in the WDM regime. However, it is encouraging to note that two of the more fundamental methods, viz., detailed configuration models and DFT estimates of <Z> converge towards agreement.

**Fig. 15**

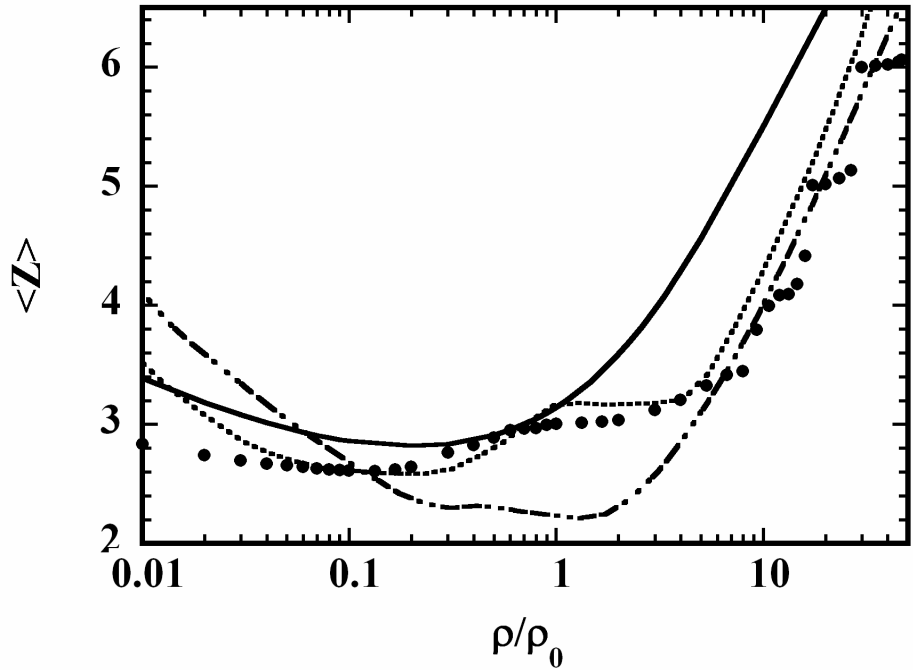

**<Z> of aluminum as a function of compression at 12.5 eV obtained from QEOS [2] (solid line), Sesame [8] (dot-dashed line), a detailed configuration model [19] (solid circles), and a neutral-psuedo-atom density-functional-theory (DFT) model [20] (dotted line). *r* is mass density and *r$_o$* is solid density of aluminum at normal conditions.**

An even more critical aspect of ionization physics is the details of ion abundance, energy and population of atomic levels. These are required for the calculation of equation of state and radiative opacities. In general, ion abundance can be revealed from the characteristics of photo-absorption lines associated with the different ions. Of particular interest here is the Al$^{+4}$ *Ka* absorption line of aluminum because of its appearance as a single, isolated line for shock compressed aluminum in the WDM regime. The use of such an absorption line as a sensitive test of <Z> has been suggested earlier [25]. A natural extension of this approach would be to test ionization models along the well-defined principal Hugoniot for shock produced Idealized Slab Plasmas. As shown in Figure 16, both the detailed configuration accounting [25] and the neutral-psuedo-atom density–functional-theory [26] models show a gradual increase in <Z> above 3 for shock pressure exceeding 20 Mbar. This can be contrasted with the much

**Fig. 16**

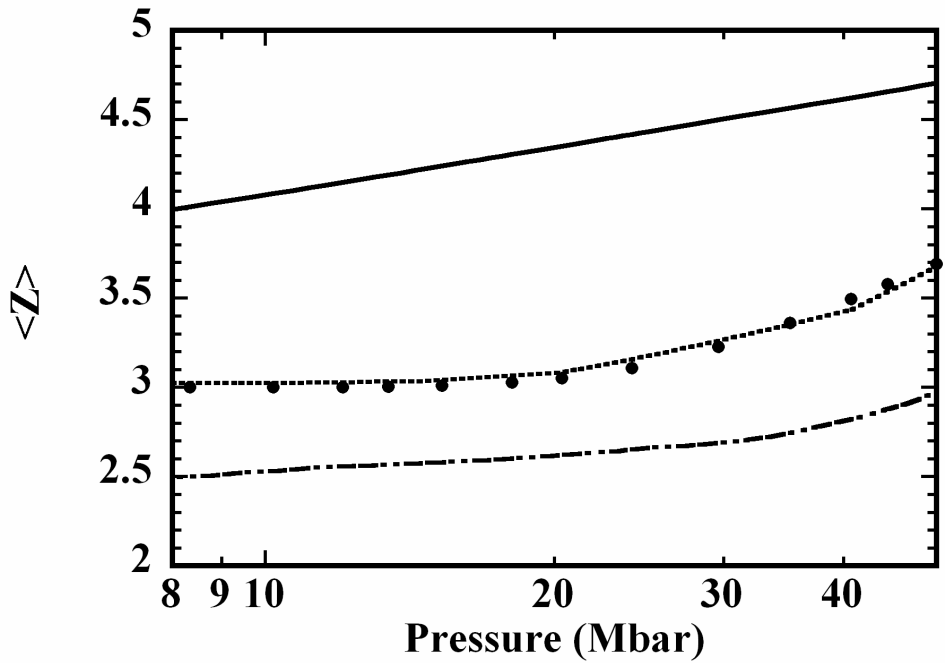

**<Z> as a function of shock pressure along the principle Hugoniot of aluminum obtained from QEOS [2] (solid line), Sesame [8] (dot-dashed line), a detailed configuration model [19] (solid circles), and a neutral-psuedo-atom density-functional-theory model [20] (dotted line).**

higher values of <Z> predicted by QEOS or the much lower values of <Z> predicted by Sesame. On the other hand, a more stringent test of theory is the ion abundance described by Figure 17(a). Specifically, the $Al^{+4}$ ion shows a rapid increase in abundance as the shock pressure varies from 10-40 Mbar. The change in abundance of such an ion can be readily measured using the opacity of its characteristics absorption line.

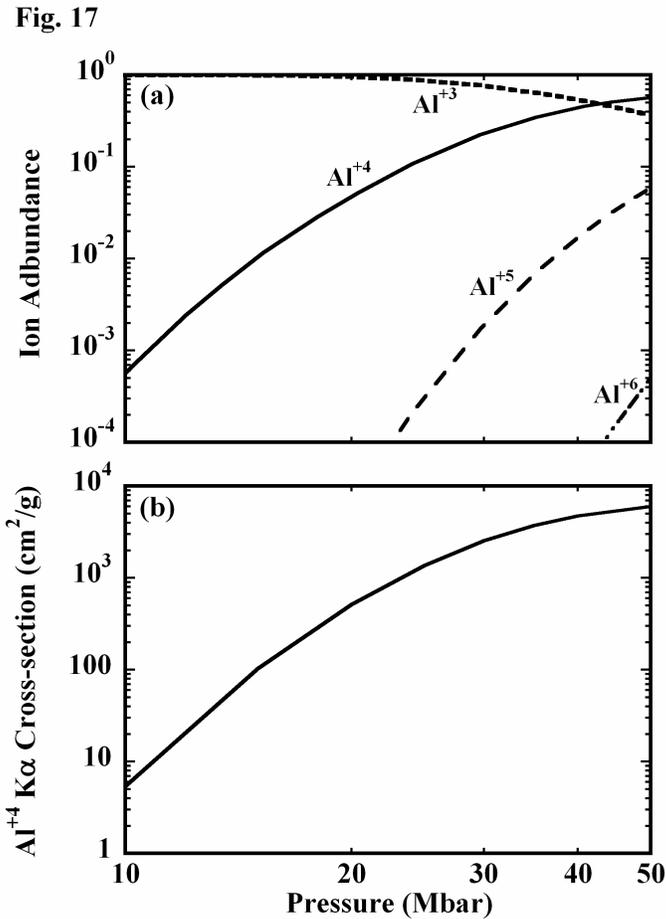

**(a) Ion abundance and (b) cross-section of the $Al^{+4}$ *Kα* absorption line as a function of shock pressure along the principal Hugoniot of aluminum.**

To compute the cross-section of the $Al^{+4}$ *Kα* absorption line, we employ oscillator strengths derived from dense plasma estimates [27] and line widths due to electron collision broadening using impact approximation [28]. Results of our calculation along the Hugoniot are presented in Figure 17(b). Variations in the absorption cross-section are dominated by changes in the $Al^{+4}$ abundance.

For opacity measurements on shock compressed matter, the usual approach is to study a sample that is sandwiched between two tamper layers in order to mitigate its expansion. Mismatch in shock impedance between the sample and the tamper causes multiple shock reflections. Furthermore, mismatch in temperature between the different layers leads to energy transport by thermal conduction. This effect becomes more dominant with decreasing sample thickness. As a result, the state of the shocked sample lies off the principal Hugoniot and varies with time. The structure of a tamped target also prevents the direct measurements of the state of the shocked sample.

As suggested earlier [25], a steady shock offers a novel and elegant means to measure the opacity of an absorption line. Its uniqueness lies in the production of a well-defined uniform state whose thickness (and hence optical depth) increases linearly with time. For the above example of a shock launched from a silicon pusher into the aluminum sample, it can readily be shown that transmitted intensity $I_T(t)$ at frequency $\boldsymbol{n}$ of a continuum X-ray source through the aluminum layer is given by

$$\ln[I_T(t)/I_O] = -[(\boldsymbol{s_n} - \boldsymbol{s}_O)\boldsymbol{r}_O U_S t] - \boldsymbol{s}_O \boldsymbol{r}_O L \qquad (1)$$

where $I_o$ is the intensity of the backlight sources reaching the aluminum layer, $L$ the initial thickness of the aluminum layer, $\boldsymbol{r}_o$ its initial density, $\boldsymbol{s}_o$ the photo-absorption cross-section for the unperturbed solid, $\boldsymbol{s_n}$ the photo-absorption cross-section for shocked aluminum, and $U_S$ the shock speed in aluminum. Equation (1) assumes that non-uniformities at the silicon-aluminum interfaces and the shock front are negligible. Accordingly, a plot of $ln(I_T/I_o)$ versus time $t$ yields a straight line whose slope is given by ($s_n$-$s_o$)$r_oU_S$. This allows a direct measurement of ($s_n$-$s_0$), and hence $s_n$, as a function of $U_S$ and provides an assessment of ion abundance associated with the absorption line.

Figure 18 shows results of our calculations for shock pressures ranging from 10-40 Mbar. The variation in the slopes corresponds to the increase in abundance of the $Al^{+4}$ ions according to the detailed configuration accounting model [25]. Such measurements will provide a direct test of ionization physics in terms of the abundance of $Al^{+4}$ ions as a function of shock pressure, as illustrated in Figure 17(b). Evidently, the use of absorption lines as a probe requires that the lines are well resolved. Clear observations of fluorine-like and oxygen-like absorption lines in shocked aluminum have been reported [29].

**Fig. 18**

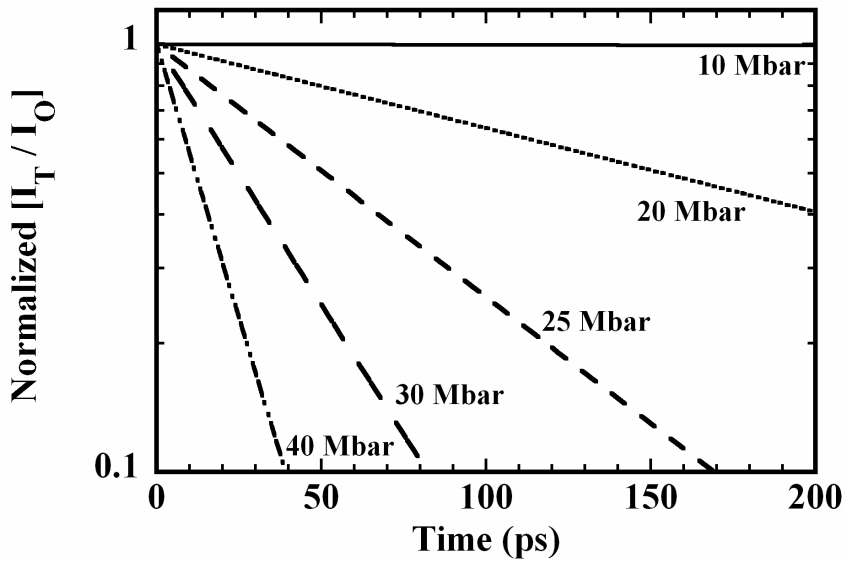

**Plot of the normalized transmission at the $Al^{+4}$ $K\alpha$ absorption line as a function of time for shock waves of different pressure.**

III.b   Effect of strong coupling on continuum states

Closely related to pressure ionization is the effect of a strongly coupled plasma environment on continuum states. A well-known manifestation of this is continuum lowering [30]. How this is related to the electron chemical potential in the plasma, and to the various many-body effects in WDM are discussed in detail by Perrot et al. in Ref. [26]. Apart from changes in ionization balance and average ionization, such an effect can be revealed by changes in photo-absorption edges [31]. Figure 19 shows a comparison of the photo-absorption cross-sections for unperturbed aluminum [32] and shocked aluminum at 25 Mbar [33]. For the cold material, the abrupt change at the *K*-shell photo-absorption edge corresponds to the sharp cutoff in the density of state in the conduction band. For the shock compressed and heated state, the photo-absorption edge becomes significantly broadened. This results from the holes generated in the conduction band by thermal excitation and changes in screening of the nucleus and bound electrons of the absorbing ions by neighboring electrons and ions.

**Fig. 19**

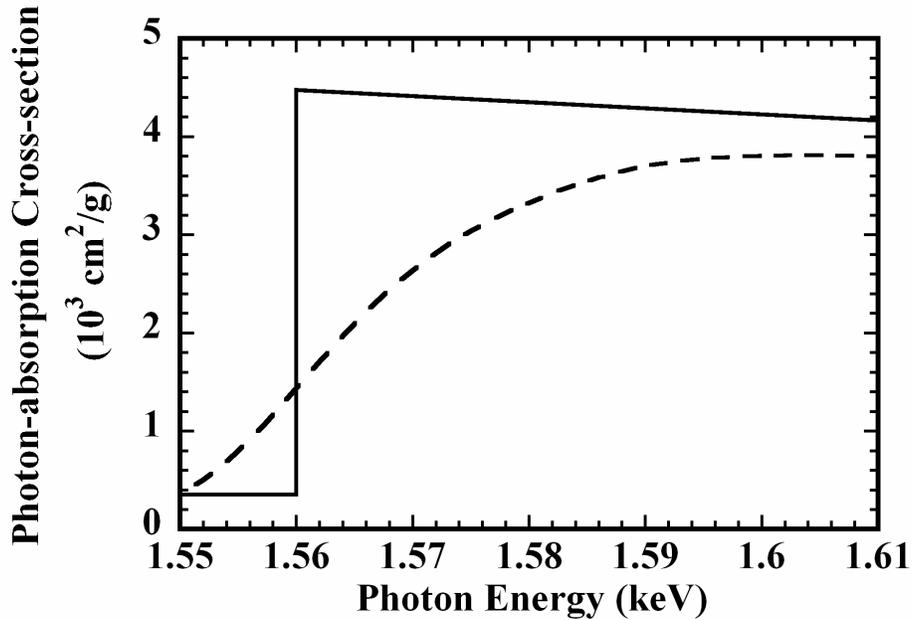

**Calculated photo-absorption cross-sections of cold aluminum (dotted line) and aluminum compressed by a shock wave at 25 Mbar (solid line).**

In experiments, the *K*-edge is usually studied from the transmission spectrum of a X-ray source through a sample. While the frequency dependence of the photo-absorption cross-section is governed only by the state of the sample, the apparent form of the transmission spectrum varies substantially with the thickness of the sample as illustrated in Figure 20. In observations using relatively thick samples and detectors with a limited dynamic range such as ultrafast streak cameras, the transmission spectrum may give the appearance of an abrupt change near the *K*-edge [31]. This has led to the misleading interpretation of an apparent shift in the edge position while there is no such identifiable shift according to the photo-absorption cross-sections. To remove such an ambiguity one

**Fig. 20**

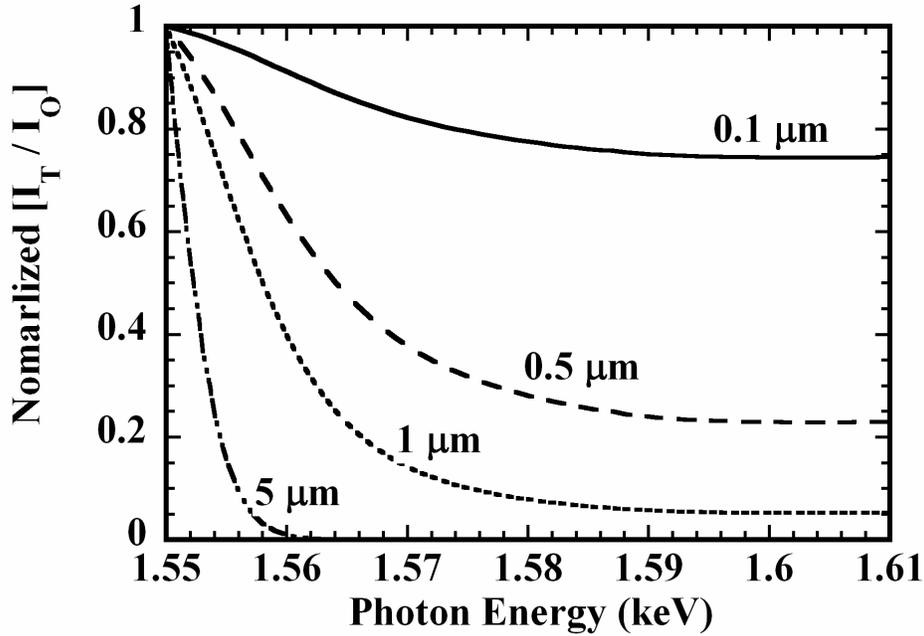

**Transmission spectra for aluminum at different shock compressed thicknesses at 25 Mbar.**

must determine the absorption cross-sections across the edge. In principle, these can be extracted from the observed transmission spectrum for a sample layer of a known thickness. For such measurements, however, it is often necessary to use tamper layers to define the extent of the sample. This approach is subject to the same difficulties discussed above.

The alternative is to adopt the use of a propagating shock wave. One option is to evaluate the cross-section from the transmission spectrum for a specific thickness of the shock compressed aluminum region. This would require accurate measurements of both shock speed and shock transit time. It would also limit the observation to a single sample thickness. The other option is to measure transmission, $I_T(t)/I_0$, at different photon frequency, **n**, for a continuum X-ray probe as a shock wave propagates in the sample, as described above for line opacity studies. As illustrated in Equation (1) the slope of $\ln[I_T(t)/I_0]$ versus $t$ will be a measure of the change in absorption cross-section, $(\mathbf{s_n}-\mathbf{s_0})$,. Results of our calculation for a 25 Mbar shock in aluminum are presented in Figure 21. The discontinuity at $h\mathbf{n}$ of 15.6 keV is due to the cold K-edge. Experimental results obtained in the form of Figure 21 will thus offer an unambiguous test of theory.

**Fig. 21**

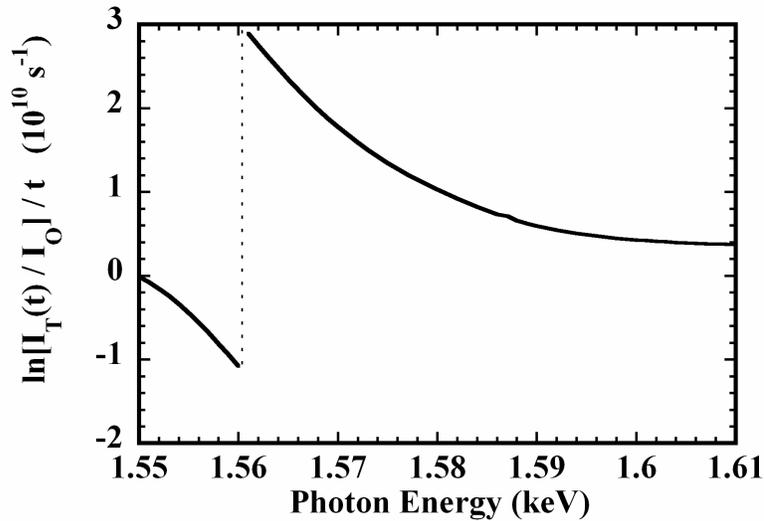

**Slope of $\ln[I_T(t)/I_0]$ versus $t$ at different X-ray energies for a 25 Mbar shock in aluminum**.

## IV    Conclusions

To meet the challenges in laboratory studies of WDM, we have presented a new approach based on the concept of Idealized Slab Plasma. This yields uniform states that can be defined by direct measurements. Two classes of Idealized Slab Plasmas are described. One of these is derived from the isochoric heating of a solid using an ultrafast energy source. The other is produced from a shock wave propagating in a solid. Specific examples are described to illustrate new means of probing some basic properties of WDM such as transport coefficient, equation of state as well as atomic physics.

In the current discussions of the isochoric heating approach, the energy sources are limited to lasers, and X-rays. However, new developments in the production of energetic electrons and ions by high-field interactions will soon render them viable for Idealized Slab Plasma experiments. In addition, one should not overlook the potential of heavy ions produced by accelerators. Although the duration of a heavy ion pulse might be limited to tens of nanoseconds, one can achieve uniform heating in sub-range samples of mm to cm scale. The relative importance of hydrodynamic expansions at the surfaces of the sample can thus become negligible in measurements of X-ray absorption cross-sections and stopping power for energetic particles.

In the discussions of Idealized Slab Plasmas produced by shock waves, we have further introduced the approach of obtaining opacity measurements along the well-defined Hugoniot. This represents an accurate and effective way to measure photoabsorption cross-sections of states that can be readily characterized by the determination of shock speeds.

Although the Idealized Slab Plasma concept is developed specifically for laboratory studies of WDM, it is an equally important concept in experimental plasma physics. It represents a unique opportunity to isolate gradient effects in the measurement of physical properties of high energy density plasmas.

Recently, the application of X-ray free electron lasers to WDM studies has been discussed [8]. New observations using ion beams have also been reported [34-37].

**Acknowledgment**

This work is supported by the Natural Sciences & Engineering Research Council of Canada and the U.S. Department of Energy (LLNL Contract No. W-7405-ENG-48).